\documentclass[preprint,12pt,authoryear]{elsarticle}

\usepackage{amssymb}
\usepackage{amsmath}
\usepackage{subcaption}
\usepackage{slashed}
\usepackage[colorlinks=true,
            linkcolor=blue,
            urlcolor=blue,
            citecolor=blue]{hyperref}

\newcommand{\calj}{\mathcal J}
\newcommand{\bfq}{\mathbf{q}}
\newcommand{\nbfq}{|\mathbf{q}|}
\newcommand{\bfx}{\mathbf{x}}
\newcommand{\nbfx}{|\mathbf{x}|}
\newcommand{\bfcalj}{\boldsymbol{\mathcal J}}
\newcommand{\bfp}{\boldsymbol{p}}

\def \a {\alpha}
\def \b {\beta}

\def \e {\epsilon}
\def \k {\kappa}
\def \l {\lambda}
\def \m {\mu}
\def \n {\nu}
\def \r {\rho}
\def \s {\sigma}

\journal{Nuclear Physics B}

\begin{document}

\begin{frontmatter}



\title{Electron Energy Loss Spectroscopy of oriented targets and magnetic transitions} 


\author{Ioannis Iatrakis} 

\affiliation{organization={ThermoFisher Scientific},
            addressline={Achtseweg Noord 5}, 
            city={Eindhoven},
            postcode={5651 GG},
            country={The Netherlands}}

\author{Valerii Brudanin} 

\affiliation{organization={Delft University of Technology},
            addressline={, Mekelweg 5}, 
            city={Delft},
            postcode={2628 CD}, 
            country={The Netherlands}}

\begin{abstract}
Electron beam energies in Transmission Electron Microscopes (TEMs) reach the relativistic realm constituting Quantum Electrodynamics (QED) the appropriate framework for the study of electron matter interaction in TEMs. We focus on the inelastic scattering of relativistic electrons from a generic oriented target. The inelastic differential cross section factorizes to the fast electron part which is calculated analytically, and the dynamic form factor of the target, which encodes the response of the medium to the interaction with the beam. The properties of the dynamic form factor of oriented targets are analyzed. We then derive the scattering cross section of electrons by magnetic targets where spin-flip transitions are induced. We comment on the kinematic regimes where the coefficient of the transverse magnetic interaction is amplified compared to the coulomb matrix element.
\end{abstract}




\end{frontmatter}



\section{Introduction}
\label{sec:intro}

The relativistic nature of electron dynamics in a TEM has a significant influence on electron energy-loss spectra (EELS). In addition to the kinematic effects on scattering, the interaction mediated by virtual photon exchange plays a central role in shaping the inelastic differential cross section \cite{PhysRevB.72.045142, sorini2008, iatrakis2025manybodyqedeffectselectronatom}. In particular, transverse photon exchange is also responsible for enabling spin-flip excitations in magnetic materials.

We work out the differential cross section of an electron beam from a generic target which interacts electromagnetically with the beam in the context of Quantum Electrodynamics (QED) to its full generality. The differential cross section is expressed in terms of the relativistic dynamic form factor. The form factor is written as the Fourier transform of the electromagnetic two-point function which is related to the retarded Green function through the well-known dispersion relations, \cite{Weinberg_1995}, \cite{10.21468/SciPostPhys.10.2.031}. As a result, transport coefficients such as the conductivity and magnetic susceptibility can be connected to the measured current two-point function using the fluctuation dissipation theorem.

We further focus on the study of the electron inelastic scattering from an oriented target. It is well known that for anisotropic materials EELS results depend on the orientation of the sample, \cite{PhysRevB.59.12807}. We study how the longitudinal and transverse parts of the interaction depend on the orientation of the target in a generic fashion. Semi-relativistic corrections have been proven important on the explanation of the magic angle in EELS experiments in the past,
\cite{PhysRevB.72.045142, 10.21468/SciPostPhys.10.2.031}.

The framework of QED is applied to a detailed analysis of magnon excitations in the sample probed by the electron beam. Spin-flip transitions have been studied in the context of resonant X-ray spectroscopy, \cite{PhysRevB.57.14584}. EELS has been recently employed for the direct detection of magnon excitations, \cite{kepaptsoglou2025magnonspectroscopyelectronmicroscope}. The inelastic differential cross section of the electrons from a target contains contributions the longitudinal (Coulomb), the transverse electric and magnetic matrix elements corresponding to different interaction channels. The interaction which results in a spin-flip of the target is induced by the magnetic virtual photons of all possible angular momenta, with the the magnetic dipole transitions being the dominant for low-momentum transfer. We derive the general result for the magnetic transitions and take the long wavelength limit corresponding to low momentum transfer to find the magnetic dipole transition matrix elements. We investigate how the experimental conditions impact the relative amplitude of the longitudinal versus the transverse matrix elements, which may serve as a proposal for future experiments. Specifically, we propose measuring EELS excitations at constant momentum transfer while varying the scattering angle. The resulting shape of the differential cross section as a function of the scattering angle can determine the transverse or longitudinal nature of the excitation.

\section{Quantum Electrodynamics}
\label{sec:QED}
The description of electrons and their interactions in the context of quantum mechanics led to the development of quantum field theory. The existence of the massless photon as the mediator of the electromagnetic interaction and its transformation properties under the Lorentz group leads to  gauge invariance  \cite{Weinberg_1995}. The QED lagrangian contains the photon and Dirac spinor kinetic terms and their coupling 

\begin{eqnarray}
    {\cal L}_{QED} =  - \frac{1}{4} F^{\mu\nu}(x) F_{\m\n}(x)
- \overline{\Psi}(x) \left( [\slashed{\partial} +i e \slashed{A}(x) ]+ m \right) \Psi(x) 
\label{QEDLagrangian}
\end{eqnarray}
We use the mostly plus signature for the Minkowski metric, $\eta_{\mu\nu} = \textrm{diag}(-1,1,1,1)$. The gamma matrices follow the Weyl notation, see \ref{gamma}. The photon field strength in terms of the electromagnetic potential reads $F_{\mu\nu} = \partial_\mu A_\n - \partial_n A_\m$, and $\Psi$ is the electron spinor field. The standard notation is adopted, $\slashed{S} = \gamma^\m S_\m$ and $\partial_\m = \frac{\partial}{\partial x_\mu}$. We adopt the natural units, where $\hbar = c = 1$.

The above Lagrangian is invariant under the gauge transformation 

\begin{align*}
    \delta \Psi(x) &= i \epsilon(x) \, e\, \Psi(x) \\
    \delta A_{\mu}(x) &= \partial_\mu \epsilon(x)
\end{align*}
implies the conservation of the current $\partial_\mu J_e^\mu(x) = -i e \partial_{\mu} \left( \overline{\Psi}(x) \gamma^\mu \Psi(x) \right) =  0$. 

In TEMs, the scattering of a fast electron beam from a heavy target is probed. The interaction of the two is mediated by the photon, $A_\mu(x)$, which couples to the target electromagnetic current through the standard minimal coupling

\begin{equation}
    {\cal L}_{\calj} = {\calj_\mu}(x) A^{\mu}(x)
    \label{externalsource}
\end{equation}
where $\calj^\mu(x) = (\rho(x), {\boldsymbol  \calj}(x))$ is conserved $\partial_\mu \calj^\mu = 0$. The external current operator represents the electromagnetic current of a heavy source, such as a charge and current distribution within a material, an atom, or a nucleus, whose dynamics can be described independently.

\subsection{Inelastic electron scattering}
\label{scattering}

\begin{figure}[t]
\centering
\includegraphics{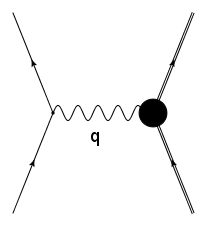}
\caption{Feynman diagram for the scattering of a high-energy electron from a heavy target.}
\label{FD}
\end{figure}

We consider the inelastic scattering of a relativistic electron of four-momentum, $p_i = (E_i, {\mathbf p_i})$, being scattered from a target at initial many-body state $| \alpha \rangle$ to final momentum, $p_f = (E_f, \bfp_f)$, and final target state $| \beta \rangle$. The electromagnetic coupling constant, $\alpha = \frac{e^2}{4 \pi} \simeq 1/137$, is small and the differential cross section can be calculated in the Born approximation. This corresponds to the tree level diagram, Fig. (\ref{FD}).

The differential cross section is written in terms of the S-matrix. A discussion of a formalism for systems involving bound states is presented in \cite{Gell-Mann:1953dcn}. The ionization cross section has been calculated in \cite{iatrakis2025manybodyqedeffectselectronatom}

\begin{equation}
d \sigma(p_f,\beta ; p_i, \alpha) =  \frac{\alpha}{2 \pi} \frac{| \mathbf p_f |}{|\mathbf p_i|} \frac{d E_f d \Omega_f}{q^4}
L_{\mu \nu} W^{\mu \nu},
\label{diffcrosssectiongen}
\end{equation}
where $q^2 = - E^2 + \mathbf{q}^2$. $L_{\mu\nu}$ and $W_{\mu\nu}$  are second rank tensors under Lorentz transformations and of course the differential cross section is invariant under Lorentz transformations. 
\begin{align}
L_{\mu \nu} & = p_{i \, \mu}  \, p_{f \, \nu} +p_{i\, \nu} \,\, p_{f \, \mu} -(p_i \cdot p_f +m^2) \eta_{\mu \nu}
\label{Ltensor}\\
W_{\mu \nu} & =\sum_\beta \int d \beta \ \delta( E_i - E_f-E)  \langle \alpha | \calj_{\n}^{\dagger}(\bfq) | \beta \rangle \, \langle \beta | \calj_{\m}(\bfq) | \alpha \rangle  \, .
\label{Wtensor}
\end{align}
The free electron tensor, $L_{\mu \nu}$, is calculated from unpolarized Dirac plane wave solutions and describes the response of a relativistic electron beam to the interaction with the target. For different types of asymptotic states or a polarized beam, $L_{\mu\nu}$ can be readily calculated, since they are solutions of the non-interacting Hamiltonian.

$W_{\mu\nu}$ is more interesting since it describes the response of the target to the electromagnetic interaction with the incoming beam electrons. This is the relativistic generalization of the dynamic structure or form  factor, \cite{PhysRev.95.249}. Similar expression is derived in the context of scalar QED for EELS analysis in \cite{10.21468/SciPostPhys.10.2.031}. In \ref{app:SQED}, a detailed calculation of the cross section within scalar QED is shown and it is compared to the full answer, where the spin degrees of freedom are included. 

Taking $\calj(\bfq)$ in the Heisenberg picture and using the completeness of the final states $| \beta \rangle$, we can write the dynamic structure tensor as

\begin{equation}
W_{\mu \nu} =\int \frac{dt}{2 \pi} e^{i E t}  \langle \alpha | \calj_{\n}^{\dagger}(\bfq, t) \,  \calj_{\m}(\bfq, 0) | \alpha \rangle  \, . 
\end{equation}
The current two-point function is also related to the imaginary part of the retarded Green's function, which is defined as the spectral function of the system. In case of positive energy loss

\begin{align}
W_{\mu\nu} = -\frac{1}{\pi}\mathrm{Im} W^R_{\mu\nu}= -\frac{1}{\pi} \mathrm{Im} \sum_\beta \frac{ \langle \alpha | \calj_\mu^{\dagger}(\bfq) | \beta \rangle \langle \beta | \calj_\nu(\bfq) | \alpha \rangle }{E-E_\beta+E_\alpha +i \epsilon} \,,
\end{align}
where the Lehmann representation of the spectral function is applied for positive frequencies. Its poles and their residue indicate the energy and strength of excited states in the spectrum  revealing information about the electronic structure of the material, \cite{DeGroot2021147061}.

The two-point function of the current operator is the central quantity derived from inclusive inelastic scattering experiments of electrons by heavy targets. There has been extensive analysis of the current two-point function for different types of targets within quantum field theory, \cite{Weinberg_1995}. In electron-nucleus scattering, $W_{\mu\nu}$ is the well-known hadronic tensor. Equation (\ref{diffcrosssectiongen}) constitutes a generic expression for the EELS inelastic differential cross section, which applies to a wide range of energy loss and momentum transfer values.

\subsection{Classical electron beam}

A common theoretical approach, especially for the analysis of low energy loss experiments, is to derive the spectral function of the target sample from the EELS spectrum, \cite{RevModPhys.82.209}. This limit is particularly interesting because the transport properties of materials are commonly studied by measuring the response of the material to an external classical perturbation, which is controlled by the spectral function, \cite{10.21468/SciPostPhys.10.2.031}. 

In the limit where the incoming electron beam is considered to be a classical current, one can study the transition rate of the target under the influence of the external field. This is not the most appropriate picture for calculating the inelastic differential cross section, which corresponds to the measured spectrum of the outgoing beam electrons. However, the cross section can be expressed in terms of the spectral function of the current operator of the target, which also appears in linear response theory. Hence, the EELS spectrum can be related to measurements of the material conductivities that depict the response to an external electromagnetic field.

The assumption of a classical electron essentially means that the fermion field in the Lagrangian (\ref{QEDLagrangian}) is not dynamical, so

\begin{eqnarray}
    {\cal L}_{cl} =  - \frac{1}{4} F^{\mu\nu}(x) F_{\m\n}(x) + J_{\mathrm(cl)}^\mu A_\mu(x)  + {\cal L}_{\calj} \,.
\label{CEDLagrangian}
\end{eqnarray}
The 4-current of a classical beam electron in the limit of small momentum transfer, where quantum effects are ignored, is

\begin{equation}
 J_{\mathrm(cl)}^{\mu}(x) =  -i e \langle \bfp | \overline{\Psi}(x) \gamma^\mu \Psi(x) | \bfp  \rangle =- \frac{e}{(2\pi)^3} e^{-i q \cdot x} \frac{p^\mu}{p^0}\, ,
 \label{currentClED}
\end{equation}
This obviously does not depend on spin. The initial and final momenta are much greater that the momentum transfer $\bfp = \bfp_i \sim \bfp_f \gg \bfq$ and $p^0 = \sqrt{\bfp^2 + m^2}$. The last term in Eq. (\ref{currentClED}) is the coupling of the photon field to the atomic target, given by Eq. (\ref{externalsource}). Essentially, the dynamical field in this description is the photon field, which is sourced by two external operators: the classical electron-beam  current, $J_{\mathrm(cl)}^{\mu}$, and the target's transition current, $\calj^\mu$. 

The ground-state configuration of the above Lagrangian can be found by solving the equations of motion for the electromagnetic field, which are simply the inhomogeneous Maxwell equations. The solution with causal boundary conditions is

\begin{align}
A^{(cl)}_\mu(q) = \langle A_\mu(q)\rangle = \Delta^{\mathrm(R)}_{\mu\nu}(q) J_{\mathrm(cl)}^\nu(q)
\label{Asol}
\end{align}
where $\Delta_{\mathrm(R)}$ is the retarded photon propagator and $q = (E, \bfq)$. The vacuum expectation value of the target current under the influence of $A^{(cl)}_{\mu}$ can be calculated perturbatively to find the Kubo formula

\begin{equation}
    \langle \calj_\mu(\bfq, E ) \rangle = - W_{\mu\nu}^{\mathrm(R)}(\bfq, E) A^{\mu \, (cl)}(\bfq, E) \, ,
    \label{kubo}
\end{equation}
where causal boundary conditions are imposed, hence the  retarded Green function is applied, which is equivalent to the Feynman one in the low momentum transfer limit. The low energy limit of the retarded Green functions is related to the electromagnetic conductivity tensor, and it is determined by the dynamics of the material. The current expectation value includes information about the induced charge, current density and magnetization of the target.

\begin{figure}[ht]
\centering
\includegraphics[width=0.2\textwidth]{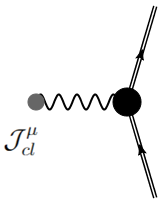}
\caption{Feynman diagram for the excitation of heavy target by an external electromagnetic field.}
\label{FDcl}
\end{figure}

In the limit of a classical electron beam, the inelastic scattering process is simplified to the Feynman diagram in Fig. (\ref{FDcl}). The corresponding S-matrix element in coordinate space reads

\begin{align}
S^{\mathrm (cl)}_{p_f,\,\beta \,;\, p_i,\,\alpha} &=  \int d^4 x_1 \int d^4x_2 \, \langle \bfp_f | J^\mu_{\mathrm (cl)} (x_1)| \bfp_i \rangle 
\, i \Delta_{\mu\nu} (x_1 - x_2)
\langle \beta | \calj^{\nu}(x_2) | \alpha \rangle\ \,.
\end{align}
Replacing the photon propagator (\ref{photonpropagator}) and the classical current from Eq. (\ref{currentClED}) we calculate the cross section

\begin{equation}
d \sigma(p_f,\beta ; p_i, \alpha) =  \frac{\alpha}{2 \pi} \frac{| \mathbf p_f | E_f}{|\mathbf p_i| E_i} \frac{d E_f d \Omega_f}{q^4}
L_{\mu \nu}^{\mathrm{(cl)}} W^{\mu \nu} \,.
\label{diffcrosssectionClED}
\end{equation}
The ratio of $E_f/E_i$ appears due to the fact that the classical beam current misses higher order $q$ corrections and depends on $E_i$ only. On the other hand, in the definition of the cross section the final asymptotic states are considered to have momentum $p_f$.  The classical electron-beam tensor reads
\begin{align}
L^{\mathrm{(cl)}}_{\mu \nu} = 2 p_{i \,\mu} p_{i \,\nu} \, ,
\label{LtensorClED}
\end{align}
which essentially agrees with the result of scalar QED which is derived in Appendix \ref{app:SQED}. This is expected due to the omission of the spin degrees of freedom in both approximations. The dynamic form tensor is still given by Eq. (\ref{Wtensor}).

\section{Oriented target}
\label{sec:poltarget}
Core loss Electron Energy Loss Spectra mainly probe atomic excitations. The atomic states are well approximated by angular momentum eigenstates. As a result, it is appropriate to expand the current 2-point function, $W_{\mu\nu}$, in a multipole series. The detailed derivation of the multipole expansion for unpolarized targets was recently presented in \cite{iatrakis2025manybodyqedeffectselectronatom}. In that work, we considered the quantization axis and the direction of $\bfq$ identical without any loss of generality, since the initial states are averaged considering an isotropic distribution. For a polarized target, this is no longer the case. However, we can calculate the transition matrix elements in a coordinate system defined by momentum transfer and then rotate the target orientation axis to an arbitrary direction. To achieve this, the current operator needs to be written in terms of the appropriate irreducible tensor operators.

Applying conservation of current to Eq. (\ref{diffcrosssectiongen})

\begin{equation}
\frac{d^2 \sigma}{dE \, d\Omega_f} =  \frac{\alpha}{2 \pi} \frac{| \boldsymbol{p}_f|}{| \boldsymbol{p}_i |} \frac{1}{q^4}
\left( 2 \calj_{\b\a}^\mu P_\mu \calj_{\b\a}^{\nu \, \dagger} P_\nu + \frac{q^2}{2} \calj_{\b\a}^{\m} \calj_{\b\a \,\m}^{\dagger} \right)
\label{diffcrosssection}
\end{equation}
where $P_\mu = \frac{p_i+p_f}{2}$ and $\calj_{\beta\alpha}^\mu =\langle \beta | \calj^\mu | \alpha \rangle$. The charge and current density operators in terms of the momentum transfer are
\begin{align}
\r(\bfq) = \int d^3x \, e^{-i \bfq \cdot \bfx} \r(\bfx)
\,,\,\,
{\boldsymbol \calj}(\bfq) = \int d^3x \,e^{-i \bfq \cdot \bfx} {\boldsymbol \calj}(\bfx) \,.
\end{align}
It is natural to expand the vector current in a coordinate system where the z-axis aligns with the momentum transfer $\bfq$

\begin{equation}
\hat{\boldsymbol e}_0= \hat{\boldsymbol e}_z= \frac{\bfq}{\nbfq} \,\,,\,
\hat{\boldsymbol e}_{\pm 1}= \mp \left( \hat{\boldsymbol e}_x \pm i \hat{\boldsymbol e}_y\right)
\label{qsystem}
\end{equation}
The conservation condition of the current, $q_\mu \calj^\mu(\bfq)=0$, constrains the longitudinal component of the 4-current to $\nbfq \, \hat{\boldsymbol e}_0 \cdot {\boldsymbol \calj}(\bfq) = q_0 \, \rho(\bfq)$. The transverse component

\begin{equation}
\bfcalj_\perp(\bfq)  =  \sum_{M=\pm 1} \calj_M(\bfq) \, \hat{\boldsymbol e}_M^\dagger \,, 
\end{equation}
where $\calj_M$ is the spherical component of the vector current.
The Wigner-Eckart theorem is applied to the  charge density matrix element, where we choose a coordinate system with $\bfx$ collinear with $\bfq$

\begin{align}
\langle \beta | \rho(\bfq)  | \alpha \rangle = \sqrt{4 \pi}
\sum_J (-i)^J  \sqrt{2 J+1} 
(-1)^{J_\b-M_\b}
\begin{pmatrix}
J_\b & J & J_\a \\
-M_\b & 0 & M_\a
\end{pmatrix}
\langle J_\beta || T_J^{\mathrm(coul)} || J_\alpha \rangle
\label{rho0ME}
\end{align}
and similarly the components of the transverse current are

\begin{align}
\langle \beta | \calj_M(\bfq)  | \alpha \rangle = -\sqrt{2 \pi} \sum_{J=1}^{\infty} & (-i)^J \sqrt{2 J +1} 
(-1)^{J_\b-M_\b}
\begin{pmatrix}
J_\b & J & J_\a \\
-M_\b & M & M_\a
\end{pmatrix}
\nonumber \\
&
\left(
\langle J_\beta || T_J^{\mathrm(el)} || J_\alpha \rangle +
M \langle J_\beta || T_J^{\mathrm(mag)} || J_\alpha \rangle
\right) \,,
\label{j0ME}
\end{align}
where

\begin{align}    
T_J^{M\, \mathrm(coul)}  =& \int d^3 \bfx \, j_J\left(\nbfq \nbfx\right) Y_J^M(\hat{\bfx}) \rho(\bfx)
\nonumber \\
T_J^{M \, \mathrm(mag)} =&  \int d^3 \bfx \, j_J(\nbfq \nbfx) \mathbf{Y}_{JJ}^M(\hat{\bfx}) \cdot{\boldsymbol \calj}(\bfx)
\label{tmatrixJ}
\\ 
T_J^{M \, \mathrm(el)} =&  \int d^3 \bfx \, 
\left[ 
\sqrt{\frac{J+1}{2 J +1}} j_{J-1} \left(\nbfq \nbfx\right) \mathbf{Y}_{J \, J-1}^M(\hat{\bfx}) \cdot {\boldsymbol \calj}(\bfx) 
\right. \nonumber \\ & \left. 
-\sqrt{\frac{J}{2 J +1}} 
j_{J+1}\left(\nbfq \nbfx\right) \mathbf{Y}_{J \, J+1}^M(\hat{\bfx})\cdot {\boldsymbol \calj}(\bfx)
\right].
\nonumber
\end{align}
The detailed derivation of irreducible spherical tensor operators in Eq.(\ref{tmatrixJ}) is shown in \cite{iatrakis2025manybodyqedeffectselectronatom}. We explicitly derive the reduced magnetic matrix element for a relativistic target which is described by solutions of the Dirac equation, in Appendix \ref{multipoleappendix}. The spin-flip part of the magnetic matrix element is further explored in Section \ref{sec::magtransitions}.

The matrix elements can now be rotated to the fixed coordinate system of the target atom, $O'(x',y',z')$. The momentum transfer with respect to $O'$ is $\hat \bfq =  \cos \phi' \cos \theta' \hat{\boldsymbol e}_{x'}+ \cos \phi' \sin \theta' \, \hat{\boldsymbol e}_{y'} + \cos \theta' \hat{\boldsymbol e}_{z'}$. The target wavefunction can be rotated to the $O'$ system by the rotation operator

\begin{equation}
D(R') =D_{\hat{\boldsymbol e}_{z}}(\phi') D_{\hat{\boldsymbol e}_{y}}(\theta') D_{\hat{\boldsymbol e}_{z}}(-\phi')\, .
\end{equation}

where $D_{\hat{\boldsymbol e}}(\phi)=exp(-i {\boldsymbol J} \cdot {\hat{\boldsymbol e}} \, \phi)$, and ${\bf J}$ is the angular momentum operator. The rotation is parametrized by the Euler angles, $R'=\{\phi', \theta', -\phi'\}$ It is well known that a matrix element can be transformed to different coordinate systems by applying the transformation to the system states or the operator themselves. Since we wrote the current operator in terms of irreducible tensor operators of definite angular momentum, it is natural to rotate the operators appropriately from the coordinate system (\ref{qsystem}) to $O'$ using their defining property

\begin{equation}
D^\dagger(R') T_J^M D(R') = \sum_{M'=-J}^J D_{MM'}^{J}(R')^* T_J^{M'}  \,.
\end{equation}
Eqs. (\ref{rho0ME}) and (\ref{j0ME}) become

\begin{align}
\langle \beta | \rho(\bfq)  | \alpha \rangle = 
4 \pi \sum_{J=0}^{\infty} 
& (-i)^J  (-1)^{J_\b - M_\b}
\begin{pmatrix}
J_\b & J & J_\a \\
-M_\b & M' & M_\a
\end{pmatrix}
\nonumber \\
& \times D_{0 M'}^J (R')^*
\langle J_\beta || T_J^{\mathrm(coul)} || J_\alpha \rangle
\label{rho0R}
\end{align}
and the components of the transverse current are

\begin{align}
\langle \beta | \calj_{M}(\bfq)  | \alpha \rangle = -\sqrt{2 \pi} \sum_{J=1}^{\infty} (-i)^J \sqrt{2 J +1} 
(-1)^{J_\beta - M_\beta}
\begin{pmatrix}
J_\b & J & J_\a \\
-M_\b & M' & M_\a
\end{pmatrix}
\nonumber \\
 \times D_{MM'}^J(R')^*
\left(
\langle J_\beta || T_J^{\mathrm(el)} || J_\alpha \rangle +
M \langle J_\beta || T_J^{\mathrm(mag)} || J_\alpha \rangle
\right) \,.
\label{j0R}
\end{align}
 Those are the most general expressions for the transition matrix elements in QED, for a target of definite orientation and any angular momentum. The transition operator obviously includes all the multipole contributions and is not restricted to the dipole one. The EELS spectrum is given by the inelastic differential cross section which depends on the norm of the transition matrix elements. In the case of an oriented atom, the different matrix elements, i.e. Coulomb, transverse electric, and magnetic mix. The generic mixed product of two such matrix elements with the initial state ensemble described by a generic density matrix $\langle M'_\a | \rho_i | M_\a \rangle$ reads:

\begin{align}
\langle \beta | {\cal O}_{M_1}^{\mathrm(i_1)} | \alpha \rangle^* \langle \beta | {\cal O}_{M_2}^{\mathrm(i_2)} | \alpha \rangle =& 
\sum_{M_\a ,M_\a'}
\sum_{J, J'} C_J C_{J'} \langle M'_\a | \rho_i | M_\a \rangle
D_{M_{1}M}^J(R')
 D_{M_{2}M'}^{J'}(R')^*
 \nonumber \\
&
\begin{pmatrix}
J_\b & J & J_\a \\
-M_\b & M & M_\a'
\end{pmatrix}
\begin{pmatrix}
J_\b & J' & J_\a \\
-M_\b & M' & M_\a
\end{pmatrix}
\nonumber \\
&
\langle J_\beta || T_J^{\mathrm(i_1)} || J_\alpha \rangle^* \langle J_\beta || T_{J'}^{\mathrm(i_2)} || J_\alpha \rangle
\end{align}
where $\mathrm{i_1, i_2 = coul, el, mag}$, denote any of the above matrix elements. $\langle M_a' | \rho_i | M_a \rangle$ is density operator of the initial target state. The product of the two rotation operators gives

\begin{align}
D_{M_{1}M}^J(R') D_{M_{2}M'}^{J'}(R')^*
=\sum_{j, m, m'} 
& 
(2 j + 1) (-1)^{M_2 - M'}
\begin{pmatrix}
J & J' & j \\
M_1 & -M_2 & m'
\end{pmatrix}
\nonumber \\
& 
\begin{pmatrix}
J & J' & j \\
M & -M' & m
\end{pmatrix}
D_{m' m}^j(R')^*
\end{align}
leading to

\begin{align}
\langle \beta | {\cal O}_{M_1}^{\mathrm(i_1)} | \alpha \rangle^* \langle \beta | {\cal O}_{M_2}^{\mathrm(i_2)} | \alpha \rangle =&
\sum_{M_\a ,M_\a'}
\sum_{J, J'} {\tilde C}_J {\tilde C}_{J'} (-1)^{M'_\a} \langle M_\a' | \rho_i | M_\a \rangle (-1)^{-M_\b-M'-M}
\nonumber \\
&
\sum_{j, m, m'}
D_{m' m}^j(R')^* 
\begin{pmatrix}
J & J' & j \\
M_1 & -M_2 & m'
\end{pmatrix}
\begin{pmatrix}
J & J' & j \\
M & -M' & m
\end{pmatrix}
\nonumber \\
&
\begin{pmatrix}
J_\b & J & J_\a \\
-M_\b & M & M_\a'
\end{pmatrix}
\begin{pmatrix}
J_\b & J' & J_\a \\
-M_\b & M' & M_\a
\end{pmatrix}
\nonumber \\
&
\langle J_\beta || T_J^{\mathrm(i_1)} || J_\alpha \rangle^* \langle J_\beta || T_{J'}^{\mathrm(i_2)} || J_\alpha \rangle
\end{align}
The sum over the Wigner-3j symbols can be simplified by the defining property of the 6j symbols, \cite{edmonds1996angular},

\begin{align}
\sum_{M_\b , M', M} 
&
(-1)^{-M_\b-M'-M} 
\begin{pmatrix}
J_\b & J & J_\a  \\
-M_\b & M & M_\a' 
\end{pmatrix}
\begin{pmatrix}
J_\b & J' & J_\a \\
-M_\b & M' & M_\a
\end{pmatrix}
\begin{pmatrix}
J & J' & j \\
M & -M' & m
\end{pmatrix}
\nonumber \\
=&
(-1)^{2 J'+J_\beta + 2 J_\alpha + j}
\begin{pmatrix}
J_\a & j & J_\a \\
M_\a' & -m & -M_\a
\end{pmatrix} 
\begin{Bmatrix}
J_\a & j & J_\a \\
J' & J_\b & J 
\end{Bmatrix} \,.
\end{align}
As a result, the product of any pair of matrix elements (longitudinal, electric or magnetic) reads

\begin{align}
\langle \beta | {\cal O}_{M_1}^{\mathrm(i_1)} | \alpha \rangle^* \langle \beta | {\cal O}_{M_2}^{\mathrm(i_2)} | \alpha \rangle =&
\sum_{M_\a ,M_\a'}
\sum_{J, J'} {\tilde C}_J {\tilde C}_{J'} (-1)^{M'_\a} \langle M_\a' | \rho_i | M_\a \rangle
\nonumber \\
&
\sum_{j, m, m'}
D_{m' m}^j(R')^* 
\begin{pmatrix}
J & J' & j \\
M_1 & -M_2 & m'
\end{pmatrix}
\begin{pmatrix}
J_\a & j & J_\b \\
M_\a' & -m & -M_\a
\end{pmatrix}
\nonumber \\
&
\begin{Bmatrix}
J_\a & j & J_\a \\
J' & J_\b & J 
\end{Bmatrix}
\langle J_\beta || T_J^{\mathrm(i_1)} || J_\alpha \rangle^* \langle J_\beta || T_{J'}^{\mathrm(i_2)} || J_\alpha \rangle
\end{align}
The prefactors $\tilde{C}$ depend on $J$ and are straightforward to find, depending on the type of operator. We will show a concrete example where those are worked out later in this section.
This is a generic expression for all the mixing terms appearing in the norm of the transition matrix element where the coulomb and transverse parts mix. The angular integrals calculated analytically in terms of the 6j symbols which are readily computed. The reduced matrix elements include only radial integrals which depend on the initial and final target states. Hence this is a generic result which can be used for any type of model of the material and either low loss or core loss excitations of any angular momentum.

In the low momentum transfer limit the cross section is dominated by the coulomb and electric dipole matrix element. The transverse electric part is proportional to the coulomb matrix element for low $\bfq$, \cite{DeForest:1966ycn},

\begin{align}
\langle \b || T_{J}^{(e)}  || \a \rangle  \simeq
\frac{E}{\nbfq}\sqrt{\frac{J+1}{J}}
\langle \b || T_{J}^{\mathrm(coul)}  || \a \rangle
\end{align}
It is then useful to show a concrete example of the ionization cross section in the limit of low q for an oriented target of a certain state $| J_\alpha M_\alpha \rangle$. In this case the density matrix is $\langle M'_\a | \rho_i | M_\a \rangle = \delta_{M'_\a M_\a}$ and the norm of the charge density operator summed over all the possible final states $|J_\b M_\b \rangle$ is

\begin{align}
\sum_{M_\b} \left| \langle \b | \r(\bfq) | \alpha \rangle \right|^2 & = 
4\pi \sum_{J \, J'\, l} (-i)^{J-J'} 
\langle \a ||T_{J'}^{\mathrm(coul) \, \dagger} || \b \rangle 
\langle \b ||T_J^{\mathrm(coul)} || \a \rangle 
\nonumber \\
& \times [J, J']^{1/2} [l] (-1)^{J'-J+J_\b+2 J_\a} (-1)^{M_\a} 
\nonumber \\
&  P_l\left( \hat{\bfq} \cdot \hat{\mathbf{e_z}} \right)
\begin{pmatrix}
J_\a & J_\a & l \\
-M_\a & M_\a & 0 
\end{pmatrix}
\begin{pmatrix}
J & J' & l\\
0 & 0 & 0 
\end{pmatrix}
\begin{Bmatrix}
J_\a & l & J_\a \\
J' & J_\b & J 
\end{Bmatrix}
\end{align}
where $[x,y,\ldots] = (2x+1)(2y+1)\ldots$. The ionization cross section is then given by

\begin{align}
\frac{d^2 \sigma}{d\Omega_f dE} &= \frac{e^2}{2 (2 \pi)^2} \frac{| \boldsymbol{p}_f|}{| \boldsymbol{p}_i|}\,
\frac{1}{q^4}
C_L(q) 
\sum_{M_\b} \left| \langle \b | \r(\bfq) |\alpha \rangle \right|^2 \,.
\label{QEDDCSrho}
\end{align}

In case of a fully isotropic target, we average over all initial states with a diagonal density matrix ($\langle M_\a | \rho_i | M_\a \rangle = \frac{1}{2 J_\a + 1}$) and the norm of the charge density matrix operator is equal to the norm of the reduced matrix of $T_J^{\mathrm (coul)}$ up to a constant factor. The same is true for the transverse matrix elements, and the full cross section reduces to 

\begin{align}
\frac{d^2 \sigma}{d\Omega_f dE} &= \frac{e^2}{2 \pi} \frac{p_f}{p_i}\,
\frac{1}{q^4} \frac{1}{2 J_\a +1} 
\left[
C_L(q) \, \sum_{J=0}^{\infty} 
\left| \langle \b ||T_J^{\mathrm(coul)} || \a \rangle  \right|^2 
\right .
\nonumber \\
& \left .
+ C_T(q) \sum_{J=1}^{\infty} 
\left(
\left| \langle \b || T_J^{\mathrm(el)} || \a \rangle \right|^2  + \left| \langle \b || T_J^{\mathrm(mag)} || \a \rangle \right|^2
\right) \right] \,.
\label{QEDDCSFinal}
\end{align}
The detailed derivation for unoriented targets has been shown in \cite{iatrakis2025manybodyqedeffectselectronatom} . The coefficients $C_L$ and $C_T$ are defined in Eq.(\ref{CTCL}).

\section{Magnetic transitions}
\label{sec::magtransitions}
The experimental quest for magnons in EELS has been an active field of research in the TEM community, \cite{kepaptsoglou2025magnonspectroscopyelectronmicroscope}. Magnon excitation by an electron beam has been theoretically studied in the low energy limit, \cite{MENDIS2022113548, nascimento2024theorymomentumresolvedmagnonelectron}. The interaction, which triggers spin-flip transitions, is due to the exchange of transverse photons between the incident electron and the material. This can be well described in the context of Quantum Electrodynamics. This interaction is already contained in the magnetic matrix element of the result (\ref{tmatrixJ}). We will here show how it asymptotes to the known low energy interaction and derive the cross section due to the magnetic moment part of the electromagnetic current. We also explore how the electric and magnetic contributions depend on the kinematics of the experiment and suggest possible ways for amplification of the magnon excitations by direct interaction with an electron beam.

Applying the property of the vector spherical harmonic
\begin{equation}
\mathbf{Y}_{JJ}^M(\hat{\bfx}) = \frac{1}{\sqrt{J(J+1)}}
\boldsymbol{L} Y_{J}^M(\hat{\bfx})
\end{equation}
The magnetic part of the transition matrix, Eq. (\ref{tmatrixJ}), is

\begin{align}
\langle \beta | T_J^{M \, \mathrm(mag)} | \alpha \rangle 
& = - \frac{i}{\sqrt{J(J+1)}}  \int d^3 \bfx \, 
(\bfx \times {\boldsymbol \calj}_{\beta\alpha}(\bfx))
\cdot \boldsymbol{\nabla} \left( j_J(\nbfq \nbfx) Y_{J}^M(\hat{\bfx}) \right)
\end{align}
which is the relativistic expression for the transition from $|\alpha \rangle$ to $| \beta\rangle$ through a magnetic photon. 

The current of a distribution of charges is defined as the convection current plus the curl of the magnetic moment of the distribution $\langle \bfcalj(\bfx) \rangle =\langle \boldsymbol{J}(\bfx) \rangle + \boldsymbol{\nabla} \times \langle \boldsymbol{\mu}(\bfx) \rangle$. In case of a system of particles, the two contributions can be mapped to the single particle states as in Eq. (\ref{currentdecomposition}). The matrix element is then
\begin{align}
\langle \beta | T_J^{M \, \mathrm(mag)} | \alpha \rangle
= - \frac{i}{\sqrt{J(J+1)}}  \int d^3 \bfx \, 
& \left( f^\dagger_{\beta}(\bfx){\boldsymbol L} f_{\alpha}(\bfx) +
\bfx \times \boldsymbol{\nabla} \times {\boldsymbol \mu}_{\beta\alpha}(\bfx) \right)
\nonumber \\ &
\cdot \boldsymbol{\nabla} \left( j_J(\nbfq \nbfx) Y_{J}^M(\hat{\bfx}) \right)
\end{align}

In the low momentum transfer limit, the above expression reduces to the expected expression from non-relativistic quantum mechanics. The Bessel function is expanded as

\begin{align}
j_J(\nbfq \nbfx) \simeq \frac{(\nbfq\nbfx)^J}{(2 J +1)!!}
\end{align}
and the electromagnetic current operator is given by Eq. (\ref{nonrelemcurrent}). Then

\begin{align}
\langle \beta | T_{J=1}^{M \, \mathrm(mag)} | \alpha \rangle  & = \frac{e}{2 m} \frac{1}{3}\frac{\nbfq}{\sqrt{2}}  \int d^3 \bfx \, 
 f_\beta^\dagger(\bfx) \left( L^M +\sigma^M\right) f_\alpha(\bfx)
 \end{align} 
The spin term coming from the magnetic moment operator can cause a spin-flip transition which generates a magnon in the sample. 

We now derive the spin-flip differential cross section explicitly for states of any angular momentum in terms of the radial integrals of the target state. The target wavefunctions are obviously natural to be written in the LS coupling scheme, since the spin must be a good quantum number of the state. The magnetic moment of the electromagnetic current operator will act on the spin part of the target wavefunction. The relevant matrix element reads

\begin{align}
\langle \beta | T_J^{M \, \mathrm(spin)} | \alpha \rangle
 =  \int d^3 \bfx \, \boldsymbol{\nabla} \times \left( j_J(\nbfq \nbfx) \boldsymbol{Y}_{JJ}^M(\hat{\bfx}) \right)
 \cdot \boldsymbol{\mu}_{\beta\alpha}(\bfx)
 \end{align}
The identity \cite{edmonds1996angular}
\begin{align}
\boldsymbol{\nabla} \times \left( j_J(|\bfq| |\bfx|) \boldsymbol{Y}_{JJ}^M(\hat{\bfx}) \right) 
= i |\bfq|
&\left( 
\sqrt{\frac{J}{2 J +1}} j_{J-1}(|\bfq| |\bfx|) \boldsymbol{Y}_{J J-1}^M(\hat{\bfx})
\right.
\nonumber \\
&\left.
-\sqrt{\frac{J+1}{2 J +1}} j_{J+1}(|\bfq| |\bfx|) \boldsymbol{Y}_{J J+1}^M(\hat{\bfx})
\right)
\end{align}
leads to

\begin{align}
\langle \beta | T_J^{M \, \mathrm(spin)} | \alpha \rangle
& =  
 i \nbfq  \sum_{L= J\pm1} 
\sqrt{\frac{L}{2 J +1}}
 \int d^3 \bfx \, 
 j_{L}(|\bfq| |\bfx|) \boldsymbol{Y}_{J L}^M(\hat{\bfx}) \cdot \boldsymbol{\mu}_{\beta\alpha}(\bfx)
\nonumber \\
& =
 i \nbfq  \sum_{L= J\pm1} 
\sqrt{\frac{L}{2 J +1}}
\int d^3 \bfx \, j_L\left(\nbfq \nbfx \right)
f_\b^\dagger(\bfx) \boldsymbol{Y}_{JL}^{M} \cdot \boldsymbol{\sigma} f_\a(\bfx)
\end{align}
Using the Wigner-Eckart theorem, the reduced matrix element is written in terms of the target angular and radial wavefunction 
\begin{align}
\langle \beta || T_J^{M \, \mathrm(spin)} || \alpha \rangle
=&  
 i \nbfq  \sum_{L= J\pm1} 
\sqrt{\frac{L}{2 J +1}}
\langle J_\beta ;  L_\beta \frac{1}{2} || \boldsymbol{Y}_{JL}^{M} \cdot \boldsymbol{\sigma} || J_\alpha ; L_\alpha \frac{1}{2} \rangle
\nonumber \\
&
\times \int dr \, r^2 j_L\left(\nbfq r \right)
F_\b^\dagger(r) F_\a(r)
\end{align}
where the angular integral is computed by decoupling the spherical tensors

\begin{align}
\langle J_\beta ; L_\beta \frac{1}{2} || \boldsymbol{Y}_{JL}^{M} \cdot \boldsymbol{\sigma} || J_\alpha ; L_\alpha \frac{1}{2} \rangle 
=& (-1)^{L_\beta} \sqrt{6} \sqrt{\frac{[L_\b, L_\a, L, J_\b, J_\a, J]}{4 \pi}}
\nonumber \\
&
\begin{Bmatrix}
L_\b & L_\a & L \\
\frac{1}{2} & \frac{1}{2} & 1  \\
J_\b  & J_\a & J
\end{Bmatrix}
\begin{pmatrix}
L_\b & L & L_\a \\
0 & 0 & 0
\end{pmatrix} \,.
\label{RMEAngular}
\end{align}

The differential cross section for the magnetic transitions due to the spin operator is
\begin{equation}
\left(\frac{d^2 \sigma}{dE \, d\Omega_f}\right)^{(spin)} =  \frac{e^2}{2 \pi} \frac{|\mathbf p_f|}{|\mathbf p_i|} \frac{1}{q^4} C_T(q) \sum_\beta | \langle \beta | T_J^{M \, \mathrm(spin)} | \alpha \rangle |^2
\label{diffcrosssectionmag}
\end{equation}
where $q^2 = -E^2 + \nbfq^2$.


\subsection{Kinematics of magnetic transitions and their detection}

\begin{figure*}[!htbp]
    \centering
    \begin{subfigure}[b]{0.48\linewidth}
            \centering
            \includegraphics[width=\linewidth]{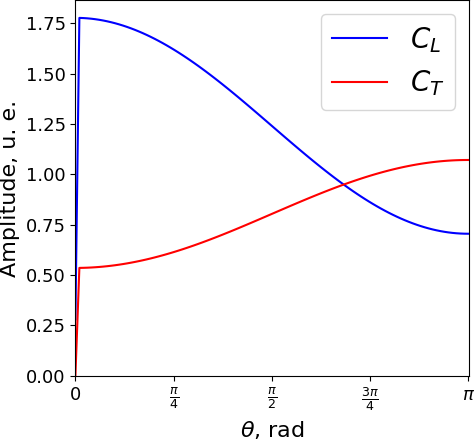}
    \caption{}
    \end{subfigure}
    \begin{subfigure}{0.5\linewidth}
            \centering \includegraphics[width=\linewidth]{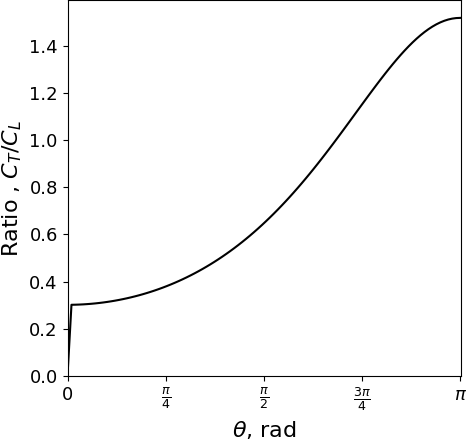}
    \caption{}
    \end{subfigure}
    \caption{The scattering coefficients as functions of the angle, for $V_{beam} = 300 kV$ and $E_{loss} = 0.1eV$. a) $C_L$ and $C_T$. c) The ratio $C_T/C_L$. The shape of the plots does not depend strongly at the value of the energy loss.}
    \label{Coefficinets2D300keV}
\end{figure*}

\begin{figure}[!htbp]
   \centering
    \begin{subfigure}[b]{0.48\linewidth}
            \centering
            \includegraphics[width=\linewidth]{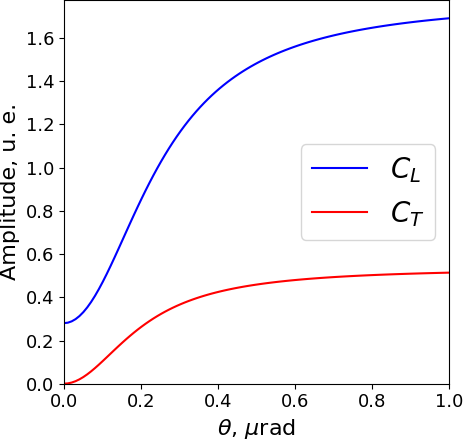}
    \caption{}
    \end{subfigure}
    \begin{subfigure}{0.5\linewidth}
            \centering \includegraphics[width=\linewidth]{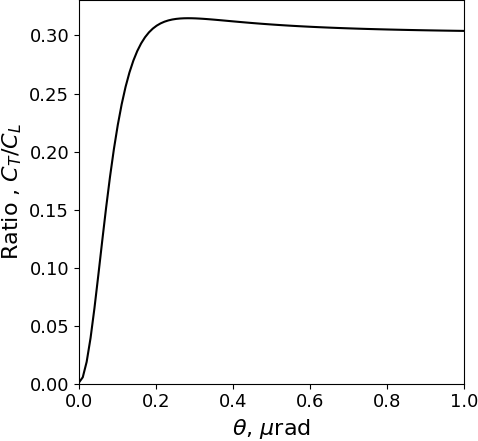}
    \caption{}
    \end{subfigure}
    \caption{The scattering coefficients as a function of the angle in the area of the steepest slope (small angles $\sim \mu rads$). For higher energy loss the steep curve extends to a few mrads. a) $C_L$ and $C_T$. c) The ratio $C_T/C_L$. $V_{beam} = 300 kV$. $E_{loss} = 0.1eV$.}
    \label{ZoomedInLowLoss}
\end{figure}

\begin{figure*}[!htbp] 
    \centering
    \begin{subfigure}[b]{0.48\linewidth}
        \centering
        \includegraphics[width=\linewidth]{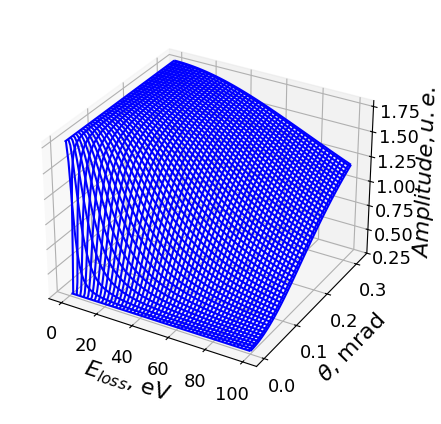}
        \caption{}
    \end{subfigure}
    \begin{subfigure}[b]{0.48\linewidth}
        \centering
        \includegraphics[width=\linewidth]{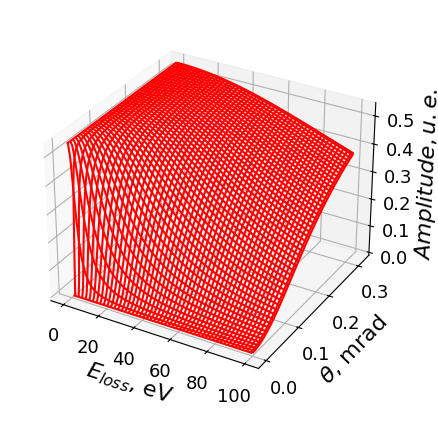}
        \caption{}
    \end{subfigure}
    \begin{subfigure}[b]{0.48\linewidth}
        \centering
        \includegraphics[width=\linewidth]{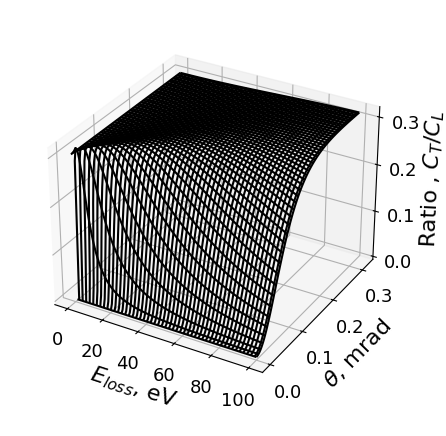}
        \caption{} 
    \end{subfigure}
    \caption{The scattering coefficients as functions of energy loss and angle. a) $C_L$. b) $C_T$. c) The ratio $C_T/C_L$. $E_{beam} = 300 keV$ }
    \label{Coefficients3D}
\end{figure*}

The magnetic part of the cross section is subleading with respect to the Coulomb and electric matrix elements by a factor proportional to the square of the fine structure constant $\sim \mathcal{O}(Z^2\alpha^2)$, at least in the long-wavelength limit. Moreover, the electric and magnetic transverse transitions have different values of parity. Our result for the inelastic scattering of the beam by magnetic excitations and, in particular, spin-flip transitions implies that it can be amplified based on the geometry of the kinematics of the scattering process. 
The longitudinal and transverse coefficients read

\begin{align}
C_L(q) &= \frac{1}{2} \left( 1- \frac{E^2}{\nbfq^2} \right)^2 \left(4 E_i^2-4 E_i \, E + E^2-\nbfq^2\right) 
\\
C_T(q) 
&=\frac{2 |\bfp_i | |\bfp_f| \, \sin^2{\frac{\theta}{2}}}{\nbfq^2} \left[ \left(|\bfp_i|+|\bfp_f|\right)^2 -2 |\bfp_i| | \bfp_f | \, \cos^2{\frac{\theta}{2}}\right]
\nonumber \\&
+ \frac{(| \bfp_i | - |\bfp_f|)^2-E^2}{2}
\label{CTCL}\end{align}
For low values of the scattering angles, the reader notices that $C_T$ is much smaller than $C_L$, Figs.(\ref{ZoomedInLowLoss}, \ref{Coefficients3D}). However, $C_T$ becomes significant for higher scattering angles and larger than $C_L$ at very large angles. This suggests that the excitations, which are probed by the transverse matrix elements only, are favored at extremely large scattering angles. Fig. (\ref{Coefficinets2D300keV}), shows that $C_T$ crosses $C_L$ at about $2.15$ rad and the ratio of the transverse to longitudinal coefficients becomes almost $1.5$ for backward scattering. This is expected since scattering experiments with nuclear targets study transverse transitions at large scattering angles.

\begin{figure*}[!htbp]
    \centering \includegraphics[width=0.6\linewidth]{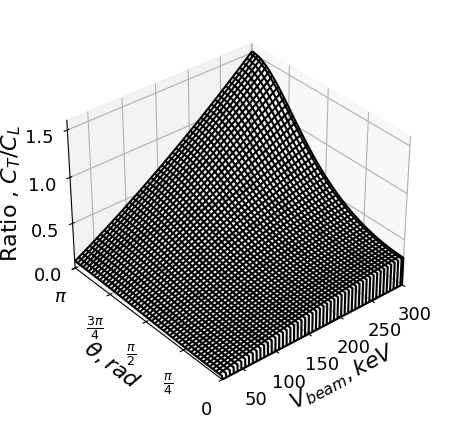}
    \caption{The ratio of scattering coefficients $C_T/C_L$ as a function of the incoming electrons energy. $E_{loss} = 0.1 eV$}
    \label{ctoverclbeamenergy}
\end{figure*}

Figure (\ref{ctoverclbeamenergy}) shows the dependence of the ratio $C_T/C_L$ on the acceleration voltage in the low–energy-loss regime. As the beam voltage increases, the ratio grows and can exceed unity. This behavior is consistent with the expectation that, in the ultra-relativistic limit, magnetic transitions become observable in backscattered electrons.

The suppression of the magnetic transition amplitude relative to the electric one indicates that direct detection of magnons in forward scattering experiments is challenging. Both the transition amplitude and the kinematic factor are significantly suppressed compared to the longitudinal channel, which dominates the scattering cross section. Our findings suggest two alternatives for detecting such signals. The straightforward detection of the magnetic energy loss signal can be enhanced in back-scattering events at high beam voltage, where relativistic effects become significant. Consequently, techniques such as reflection EELS can be effective, although the overall signal in the spectra at high beam voltage will be low, \cite{PhysRevB.82.155422}.

Since $C_L$ and $C_T$ both contribute to electric transitions, while magnetic ones involve only $C_T$, a model-independent strategy is to measure the spectrum as a function of the scattering angle while adjusting the beam voltage to keep the momentum transfer fixed, \cite{WEIGERT1964529}. In this case, the contribution from the target's current matrix elements ($T^{\mathrm{coul}}$,$T^{\mathrm{el}}$ and $T^{\mathrm{mag}}$) will be constant since they depend on $| \bfq |$. The variation in the measured EELS intensity as a function of the angle will be controlled by the coefficients $C_L$ and $C_T$. By factoring out $C_L$ the shape of the relative intensity of a transverse transition in terms of the scattering angle will be controlled by $C_T/C_L$ when the momentum transfer is constant. 

\begin{figure}[!htbp]
   \centering
    \begin{subfigure}[b]{0.42\linewidth}
            \centering
            \includegraphics[width=\linewidth]{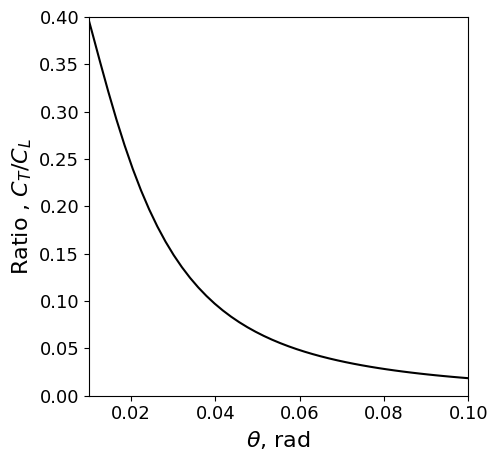}
    \caption{}
    \end{subfigure}
    \begin{subfigure}{0.41\linewidth}
            \centering \includegraphics[width=\linewidth]{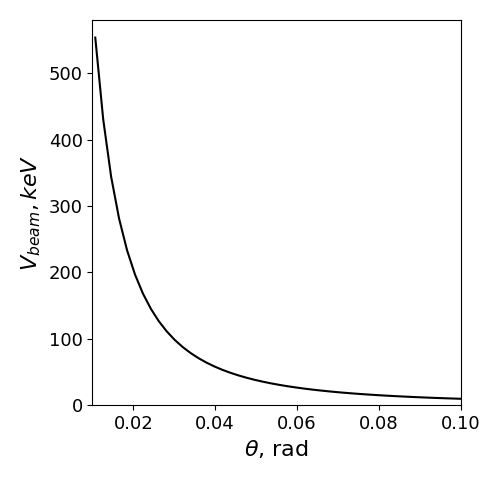}
    \caption{}
    \end{subfigure}
    \caption{(a) The ratio of the scattering coefficients as a function of the angle for constant momentum transfer, $|\bfq| = 2.6$ in atomic units.
    (b) The beam voltage as a function of the angle for constant momentum transfer $|\bfq| = 2.6$, in atomic units.}
    \label{CTCLconst}
\end{figure}

Alternative approaches to probing spin-flip transitions include resonant inelastic X-ray scattering at the core-shell edges of transition metals \cite{PhysRevB.57.14584}. This technique enables high-precision access to intermediate states in which a spin flip is induced during the resonant process. In a transmission electron microscope, an analogous strategy may be realized by measuring the wavelength-dispersive X-ray spectrum excited by a resonant electron beam, potentially providing similar insights.

The development of vortex electron beams in TEMs has led to the study of magnetic effects using electron magnetic dichroism (EMCD). In this technique, the electron probe is prepared in an angular momentum eigenstate. \cite{PhysRevB.94.174414} investigates the elastic scattering of relativistic electron beams from ferromagnetic and antiferromagnetic materials. The analysis compares beams with opposite angular momentum projections and shows that, after diffraction by the crystal, their signals differ due to interaction with the material’s induced magnetic field. We expect that the contribution for the longitudinal part has largely canceled. Notably, the resulting magnetic signal is concentrated at small scattering angles—opposite to the trend expected from the relativistic elastic scattering cross section, which increases with angle. This behavior can possibly be explained by the combined effects of the crystal geometry and the preparation of the initial vortex beam. Moreover, the magnetic signal can be further enhanced by employing beams with larger orbital angular momentum. The concept of electron magnetic dichroism in EELS was proposed in \cite{HEBERT2003463}.

\section{Summary and outlook}
EELS data are generated by the inelastic scattering of the electron beam with the material. This process is analyzed in the framework of deep inelastic scattering in Quantum Field Theory for the general type of targets and a fully relativistic electron beam,~\citet{DeForest:1966ycn, Drell:1963ej, Weinberg_1995}. We emphasize how the relativistic form factor naturally appears from the inelastic cross section calculation in QED and comment on the relation to linear response theory and the calculation of transport properties from the electromagnetic current two-point functions. Our results are generic and do not depend on the model that is used to calculate the target matrix elements. Similar construction using slightly different techniques and the model of scalar QED was presented in~\citet{10.21468/SciPostPhys.10.2.031}. 

We then bring our attention to the explicit calculation of the inelastic differential cross section for polarized targets with a specified angular momentum direction. In~\citet{iatrakis2025manybodyqedeffectselectronatom}, the ionization and excitation cross section was calculated for unpolarized atomic targets. The construction is now generalized without summing over all possible initial projections of the target angular momentum with equal weights. In this case, the cross section also depends on the quantization axis direction of the target atom. A concrete example for the Coulomb matrix element is shown for atomic states of any angular momentum, with the angular integrals being explicitly calculated in terms of the appropriate coupling coefficients.

Finally, the question of magnon excitations in EELS is investigated in the current theoretical framework. The Coulomb, electric, and magnetic contributions are all included in the current calculation and their respective contributions can be quantified on equal terms.We explicitly show that taking the long wavelength limit of the magnetic transition matrix element reproduces the known dipole magnetic moment matrix element of the target. We then calculate the differential cross section for a spin-flip transition which comes from the magnetic moment part of the electromagnetic transition current. The magnetic dipole transitions are known to have much lower intensity than the coulomb and electric dipole transitions. However, the coefficient of transverse transitions is shown to increase for large scattering angles. Our results suggest that experiments of high-energy beams at high scattering angles can show an amplification of the transverse versus the longitudinal transitions compared to their values at low scattering angles. Moreover, the coefficients $C_L$ and $C_T$, Eq.~(\ref{CTCL}), depend on the scattering angle differently, and this fact has been used in the past for the study of magnetic excitations in nuclear targets,~\citet{WEIGERT1964529}. Hence, EELS experiments with a varying scattering angle can be used to measure the dependence of transverse transitions on the angle of constant energy loss and momentum transfer. This would require the beam voltage to be modified per scattering angle. The spectrum shape in terms of the scattering angle can then reveal different types of transitions. An interesting extension of the current work is the modeling of the magnetization, charge, and current density of realistic materials, and the calculation of their EELS spectrum using the current formalism.  Another future investigation involves using the current formalism to calculate the EMCD EELS signal within the framework of QED. 

\section{ACKNOWLEDGMENTS}
We are grateful to the organizers of the "Enhanced Data Generated by Electrons (EDGE) 2024 International Workshop on Electron Energy Loss Spectroscopy and Related Techniques" for giving us the opportunity to present our work. We also thank M. Kociak for valuable discussions, and F. de Groot in particular for insightful conversations on resonant inelastic X-ray scattering and its potential analogue using an incoming electron beam. V.B. acknowledges support by an ERC Advanced Grant, under grant agreement no. 101055013.

\appendix


\section{\label{app:nonrelativistic}
Non-relativistic limit}
\subsection{\label{app:nonrelativistictarget}The target transition current}

It is common for the momentum transfer from the relativistic beam electrons to the sample to be small in EELS experiments. In this case, non-relativistic target excitations are probed. The non-relativistic limit of the full QED inelastic cross section has been derived in \cite{iatrakis2025manybodyqedeffectselectronatom} for unpolarized targets. 

Here we show how the non-relativistic expression of the transition current interacting with an external field is derived as a limit of the QED transition current.

An electron wavefunction is defined by the spinor field $\Psi(x)$ in (\ref{QEDLagrangian}), as $\psi_n(x) = \langle 0 | \Psi(x) | n \rangle$ where $n$ is the set of quantum numbers of the appropriate Hilbert space. The equation of motion of the wavefunction reads

\begin{equation}
\left(\slashed{\partial} +m + i e \slashed{A}(x) \right) \psi_n(x)=0
\end{equation}
Introducing the two-component spinor wavefunctions
\begin{equation}
\psi_n(\bfx) = 
\begin{pmatrix}
f_n(\bfx) +i \, g_n(\bfx)  \\
f_n(\bfx) -i \, g_n(\bfx)
\end{pmatrix} e^{-i E_n t}
\label{eq:diraceigen}
\end{equation}
We follow the gamma matrix definition (\ref{chiralgamma}). The equations of motion read

\begin{align}
{\boldsymbol \sigma} \cdot \left( \boldsymbol{\nabla} +e {\boldsymbol A} \right) f_n(\bfx) &= (E_n + e A^0 + m) g_n({\bfx})
\label{feom} \\
{\boldsymbol \sigma} \cdot \left( \boldsymbol{\nabla} +e {\boldsymbol A} \right) g_n(\bfx) &= -(E_n + e A^0 - m) f_n({\bfx})
\label{geom}
\end{align}
In the non-relativistic limit $E_n \simeq m+ \e_n$ where $\e_n \ll m$. We also consider the static electric field to be small compared to the rest mass of the electron. This is a reasonable assumption for the majority of excitations, which are probed in EELS experiments. Eq. (\ref{feom}) gives

\begin{equation}
g_n(\bfx) \simeq \frac{1}{2m}{\boldsymbol \sigma} \cdot \left( \boldsymbol{\nabla} +e {\boldsymbol A} \right) f_n(\bfx)
\end{equation}
The matrix element of the transition current operator in the non-relativistic limit is derived 

\begin{align}
\langle m | \bfcalj(\bfx) | n \rangle = -& i e  \overline{\psi}_m(\bfx) \gamma^\mu \psi_n(\bfx) 
\nonumber \\
= &  
i e\, \left( f_m^\dagger(\bfx) {\boldsymbol \sigma} g_n(\bfx) + g_m^\dagger(\bfx) {\boldsymbol \s} f_n(\bfx) \right)
\nonumber \\
= & \frac{i \, e}{2 m} \, \left( f_m(\bfx) {\boldsymbol \nabla} f^\dagger_n(\bfx)-  f^\dagger_m(\bfx) {\boldsymbol \nabla} f_n(\bfx) 
 +i {\boldsymbol A} f^\dagger_m(\bfx)f_n(\bfx)
 \right.
 \nonumber \\
& 
\left.
-i {\boldsymbol \nabla} \times \left( f_m(\bfx) {\boldsymbol \sigma} f_n(\bfx) \right)  \right)
\label{nonrelemcurrent}
\end{align}
This is the expression for one-particle states which can be generalized to a multi-particle fermionic state, which is generically a linear superposition of slater determinants. We notice that the spin contribution is of the same order as the convection current.  This expression can be matched with the decomposition in terms of the convection current and magnetic moment

\begin{align}
\langle \boldsymbol{J}_{mn}(\bfx) \rangle 
&= \frac{i \, e}{2 m} \, \left( f_m(\bfx) {\boldsymbol \nabla} f^\dagger_n(\bfx)-  f^\dagger_m(\bfx) {\boldsymbol \nabla} f_n(\bfx)  +i {\boldsymbol A} f^\dagger_m(\bfx)f_n(\bfx) \right)
\nonumber \\
\langle \boldsymbol{\mu}_{mn}(\bfx) \rangle
&=
\frac{e}{2 m} \,  f_m(\bfx) {\boldsymbol \sigma} f_n(\bfx)
\label{currentdecomposition}
\end{align}

\subsection{\label{app:SQED}The semi-relativistic beam and the scalar QED case}

For scattering events at low energy loss and small momentum transfer, several more approximations are common in the literature. In this regime, the contribution from the spin degrees of freedom of the probe electrons is suppressed. As a toy model, scalar QED has sometimes been employed to describe the relativistic interaction between a spinless beam and the target. However, it should be emphasized that scalar QED is not a consistent reduction of the realistic spinor QED. The two theories coincide only in the non-relativistic limit, provided that spin degrees of freedom are neglected. The Pauli approximation is a controlled low-energy expansion derived consistently from spinor QED, as shown in \ref{app:nonrelativistictarget}. 

We here calculate the cross section for scattering of a scalar beam from an external target. The scalar QED Lagrangian is

\begin{eqnarray}
    {\cal L}_{SQED} =  - \frac{1}{4} F^{\mu\nu}(x) F_{\m\n}(x)
  - D^\mu \Phi(x)^\dagger D_\mu \Phi(x)  -m^2 \Phi^\dagger \Phi+ \calj^\mu A_\mu
 \, .
\label{SQEDLagrangian}
\end{eqnarray}

The scattering of a scalar electron from four-momentum $p_i$ to $p_f$ by a heavy target is described by the differential cross section which is given by

\begin{equation}
\frac{d^2 \sigma}{dE_f \, d\Omega_f} =  (2 \pi)^4 
\frac{| \boldsymbol{p}_f|}{| \boldsymbol{p}_i |} 
E_i E_f  \, |T_{p_f,\, \beta \,;\, p_i,\, \alpha}|^2 \delta(E_i-E_f-E_\beta-E_\alpha) d\beta\,.
\label{diffcrosssectionTmatrix}
\end{equation}
The transition matrix element is calculated by the Feynman diagram in Fig. (\ref{FD}), where the external lines of the left-hand side vertex now correspond to scalar electrons.
The incoming/outgoing scalar electron plane waves contribute to the scattering amplitude a factor $\frac{e^{\pm ip x}}{(2 \pi)^{3/2}\sqrt{2 p^0}}$, where $p^0 = \sqrt{\bfp^2 +m^2}$. The photon-fermion vertex is $-e (p_f^{\mu} + p_i^\mu)$, while the coupling with the external current results in a vertex $-\calj_{\m}(x) $. The photon propagator is
\begin{align}
\Delta_{\mu\nu}(x_1,x_2)= \int \frac{d^4 q}{(2 \pi)^4} \frac{\eta_{\mu\nu}}{ q^2 - i \varepsilon} e^{i q(x_1-x_2)} \,.
\label{photonpropagator}
\end{align}
The Feynman diagram amplitude then reads

\begin{align}
T_{p_f,\, \beta \,;\, p_i,\, \alpha} &=-\frac{e}{(2 \pi)^3} \frac{p_f^{\mu} + p_i^\mu}{\sqrt{4 E_f E_i }}\frac{1}{q^2 -i \varepsilon} 
\int d^3x e^{-i \bfq \cdot \bfx} \langle \beta | \calj_{\m}(\bfx) | \alpha \rangle\ \,.
\end{align}
The cross section is calculated by the norm of $T_{p_f,\, \beta \,;\, p_i,\, \alpha}$ and it is given by Eq. (\ref{diffcrosssectiongen}), where now

\begin{align}
L^{SQED}_{\mu \nu} & = \frac{1}{2}(p_i + p_f )_\mu (p_i + \, p_f )_\nu
\end{align}
which is different by the term of $(p_i \cdot p_f + m^2) \eta_{\mu \nu}$ from the QED case. The dynamic form tensor is still given by Eq.(\ref{Wtensor}). Since $L_{\mu\nu}$ is contracted with a conserved current, $q_\mu \calj^\mu(\bfq) = 0 $, terms proportional to $q_\mu$ will not contribute to the amplitude and the SQED expression is the same as $L_{\mu\nu}^{\mathrm{cl}}$ in Eq. (\ref{LtensorClED}).

Considering the current conservation, we find the differential cross section in terms of the target's transition current
\begin{equation}
\frac{d^2 \sigma}{dE \, d\Omega_f} =  \frac{\alpha}{\pi} \frac{| \boldsymbol{p}_f|}{| \boldsymbol{p}_i |} \frac{1}{q^4}
\calj_{\b\a}^\mu P_\mu \calj_{\b\a}^{\nu \, \dagger} P_\nu \,,
\label{diffcrosssectionSQED}
\end{equation}
which is of the same form as Eq.(\ref{diffcrosssection}) with the difference of the last term $\frac{\bfq^2}{2} |\calj|^2$.

\begin{figure*}[!htbp]
    \centering \includegraphics[width=0.5\linewidth]{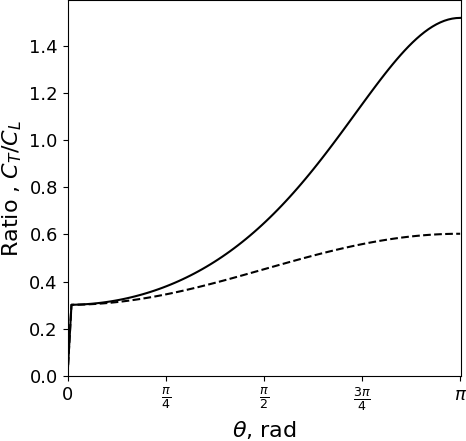}
    \caption{The ratio of scattering coefficients $C_T/C_L$ as a function of the scattering angle. Solid line: QED, Dashed line: Scalar QED $E_{loss} = 0.1 eV$}
    \label{SQEDctoverclscangle}
\end{figure*}

It is straightforward to work out the differential cross section for angular momentum eigenstates of the target to find the coefficients $C_L$ and $C_T$

\begin{align}
C^{SQED}_L(q) &= \frac{1}{2} \left( 1- \frac{E^2}{\nbfq^2} \right)^2 \left(4 E_i^2-4 E_i \, E + E^2\right) 
\\
C^{SQED}_T(q) 
&=\frac{2 |\bfp_i | |\bfp_f| \, \sin^2{\frac{\theta}{2}}}{\nbfq^2} \left[ \left(|\bfp_i|+|\bfp_f|\right)^2 -2 |\bfp_i| | \bfp_f | \, \cos^2{\frac{\theta}{2}}\right]
\label{CTCLSQED}
\end{align}
where we indeed verify that the SQED result agrees with the cross section to the lower order terms in $|\bfq|$. In  Fig. (\ref{SQEDctoverclscangle}), it is shown that the difference between the scalar QED and QED appears stronger at much higher scattering angles.

The ionization cross section has also been calculated within the framework of single-particle quantum mechanics, where the wavefunctions of the beam electrons are scattered by the electromagnetic field of the target. The corresponding effective one-particle potential can be obtained from the Fourier transform of the relevant Feynman diagram. The dominant contribution is the Coulomb interaction, while the leading relativistic correction, of order $1/c$, is given by the Breit term. A detailed derivation can be found in \cite{berestetskii1982quantum}.

\section{The multipole expansion of the magnetic matrix element}
\label{multipoleappendix}
The multipole expansion of the dynamic form tensor was analyzed in \citet{iatrakis2025manybodyqedeffectselectronatom} by coupling the appropriate tensor operators in spherical coordinates. Here we provide an alternative derivation, \cite{PhysRevA.37.307}, based on the general properties of the vector spherical harmonics as \cite{edmonds1996angular}
\begin{align}
\mathbf{Y}_{J\,J}^{M} (\hat{\bfx}) =\frac{1}{J(J+1)} \mathbf{L} Y_J^M(\hat{\bfx})
\nonumber \\
\boldsymbol{\sigma} \cdot \mathbf{L} \Omega_{\k \, m}(\hat{\bfx}) = -(1+\k)\Omega_{\k \, m}(\hat{\bfx})
\end{align}

The magnetic transition matrix can then be calculated in terms of the radial integrals over the Dirac wavefunctions of the target. Those are defined from Eq. (\ref{eq:diraceigen})
as $f_n(\bfx)=\Omega_{\k\,m}({\hat \bfx}) F_n(r)$ and $g_n(\bfx)= \Omega_{-\k\,m}({\hat \bfx}) G_n(r)$\, where the spinor spherical harmonic is

\begin{equation}
 \Omega_{\k\,m}({\hat \bfx})= 
\begin{pmatrix}
C_{j \mp \frac{1}{2}\,, \frac{1}{2}} (j,m\; m-\frac{1}{2}, \frac{1}{2}) Y^{m-\frac{1}{2}}_{j\mp \frac{1}{2}} \\
C_{j \mp \frac{1}{2}\,, \frac{1}{2}} (j,m\; m+\frac{1}{2}, \frac{1}{2})
Y^{m+\frac{1}{2}}_{j\mp \frac{1}{2}}
\end{pmatrix}
\label{sphericalspinors}
\end{equation}
and $\Omega_{-\k\,m}({\hat \bfx}) = (\boldsymbol{\sigma} \cdot \bfx)\, \Omega_{\k\,m}({\hat \bfx})$. 

The magnetic matrix element then reads
\begin{align}
\langle \b || T_{J}^{M \, (m)}  || \a \rangle &= i\, e \, 
\int d^3x j_J\left(\nbfq r\right) 
\left[ 
f_\b^\dagger(\bfx) \mathbf{Y}_{J\,J}^{M} \cdot \boldsymbol{\sigma} g_\a(\bfx)  - g_\b^\dagger(\bfx) \mathbf{Y}_{J\,J}^{M} \cdot \boldsymbol{\sigma} f_\a(\bfx)
\right] 
\nonumber \\
&= i \,e \,  
\left[
\langle \k_\b || \left[ \mathbf{Y}_{J\,J}^{M} \cdot \boldsymbol{\sigma} \right] || -\k_\a \rangle
\int dr \, r^2\, j_J\left(\nbfq r\right) F_\b^\dagger(r) G_\a (r)
\right . \nonumber \\
&- \left .
\langle -\k_\b || \left[ \mathbf{Y}_{J\,J}^{M} \cdot \boldsymbol{\sigma} \right] || \k_\a \rangle
\int dr \, r^2\, j_J\left(\nbfq r\right)  G_\b^\dagger(r) F_\a (r)
\right]
\nonumber \\
&=-i \, e \frac{(\k_\a+\k_\b)}{\sqrt{J(J+1)}}
\langle \k_\b || Y_{J} || -\k_\a \rangle
\int dr \, r^2\, j_J\left(\nbfq r\right) 
\nonumber \\
&
\times \left( F_\b^\dagger(r) G_\a(r)+G_\b^\dagger(r) F_\a (r) \right)
\end{align}
where we have defined
\begin{equation}
\langle \k_m || Y_{J} || \k_n \rangle =
\frac{(-1)^{j_m-\frac{1}{2}}}{\sqrt{4 \pi}} \, \left[ j_m, \, J, \, j_n \right]^\frac{1}{2} 
\begin{pmatrix}
j_m & J & j_n \\
\frac{1}{2} & 0 & -\frac{1}{2} \,.
\end{pmatrix}
\label{Ymatrixelement}
\end{equation}

\section{\texorpdfstring{$\gamma$}{gamma} matrix representation}
\label{gamma}
The gamma matrices satisfy the Dirac algebra 
\begin{equation}
\left\{\gamma^{\mu} , \gamma^{\nu} \right\} = 2 \eta^{\mu \nu} \,.
\end{equation}
The Dirac representation is particularly convenient for investigating the non-relativistic limit. This reads

\begin{align}
\gamma_D^0 = -i
\begin{pmatrix}
\mathbb{I}  & 0 \\
0 & -\mathbb{I} 
\end{pmatrix}
\,\,,\,\,\
\gamma_D^i = -i
\begin{pmatrix}
0 & \sigma^i  \\
-\sigma^i & 0 
\end{pmatrix}
\end{align}
The Weyl representation where the left and right handed spinors are more conveniently expressed is 

\begin{align}
\gamma_C^0 = -i
\begin{pmatrix}
0 & \mathbb{I}  \\
\mathbb{I} & 0 
\end{pmatrix}
\,\,,\,\,\
\gamma_C^i = -i
\begin{pmatrix}
0 & \sigma^i  \\
-\sigma^i & 0 
\end{pmatrix}
\,\,,\,\,\,
\beta = i \gamma^0
\label{chiralgamma}
\end{align}
A similarity transformation connects the two representations
\begin{align}
U=U^{-1} = \frac{1}{\sqrt{2}}
\begin{pmatrix}
\mathbb{I}   & \mathbb{I}  \\
\mathbb{I} & -\mathbb{I}   
\end{pmatrix} \,.
\end{align}
Then
\begin{equation}
\gamma^{\mu}_C = U \, \gamma^{\mu} U^{-1} \,\, , 
\Psi_C = U \Psi_D
\end{equation}
In both cases, we may define 
\begin{equation}
\beta = i \gamma^0 \, .
\end{equation}
Originally, Dirac's equation was written in terms of the $\alpha$ matrices defined as
\begin{align}
\alpha_4=\beta = i \gamma^0 \,\, , \,\,\, \alpha^i=i \alpha_4 \gamma^i \,.
\end{align}


\bibliographystyle{elsarticle-harv} 
\bibliography{QEDEELSrefs_Revised}

\end{document}